\documentclass[pra,showpacs,twocolumn,superscriptaddress,aps]{revtex4-1}%
\usepackage{color}
\usepackage{amsfonts}
\usepackage{amsmath}
\usepackage{amssymb}
\usepackage{mathrsfs}
\usepackage{graphicx}%

\begin{document}

\title{Entanglement generation between two spinor Bose-Einstein condensates with cavity QED}

\author{Daniel Rosseau}
\affiliation{National Institute of Informatics, 2-1-2 Hitotsubashi, Chiyoda-ku, Tokyo 101-8430, Japan}

\author{Qianqian Ha}
\affiliation{Institut polytechnique de Grenoble, Grenoble Institute of Technology, 46 avenue Félix Viallet, 38031 Grenoble Cedex 1, France}

\author{Tim Byrnes}
\affiliation{National Institute of Informatics, 2-1-2 Hitotsubashi, Chiyoda-ku, Tokyo 101-8430, Japan}

\begin{abstract}
We analyze a scheme for generating entanglement between two spinor Bose-Einstein condensates (BECs).  The BECs are off-resonantly coupled to a common photon mode and are controlled by external lasers to induce a $ S^z S^z $ interaction, where $ S^z $ is the total spin of the BEC. We directly simulate the scheme numerically for small systems and show the performance of the scheme.  The scaling of the entanglement to large scale systems under realistic conditions of spontaneous emission and cavity photon loss is analyzed. It is shown that both entanglement in the beyond-continuous variables regime can be generated, where the entanglement is of the order of the maximal entanglement between the systems. 
\end{abstract}

														
\date{\today}
\maketitle
\section{Introduction}
\label{sec:intro}

Recently there has been a large amount of interest in observing quantum mechanical phenomena on the macroscopic scale \cite{vedral08}.  The interest comes from both a fundamental and technological perspective, to both understand the quantum-to-classical transition \cite{zurek03} and develop new approaches towards quantum information processing \cite{nielsen00}. The generation of entanglement is a key target as it is an essential ingredient in many quantum information protocols, allowing for tasks which are beyond the possibility for a device based on classical physics \cite{nielsen00}.   In particular, entanglement has been generated between macroscopic systems, such as atomic ensembles \cite{julsgaard01,krauter13,bao12} and superconducting systems \cite{sillanpaa07, majer07}.  Hybrid systems interfacing two different types of systems involving macroscopic objects has also been successfully performed, such as atomic ensembles and light \cite{sherson06}, superconductors and microwaves \cite{blais04, wallraff04}, and atomic ensembles and superconductors \cite{kubo11}.  Spinor Bose-Einstein condensates (BECs), where the condensed atoms possess a spin degree of freedom, are potential candidates for the creation of non-local entanglement. By virtue of the extremely low temperatures that BECs are realized, they have been observed to have extremely long coherence times to the order of a second \cite{treutlein06}.  Recently entanglement between a BEC and a single atom was recently demonstrated \cite{lettner11}.  However, to date the entanglement of two BECs has not been demonstrated.  

\begin{figure}
   \scalebox{0.3}{\includegraphics{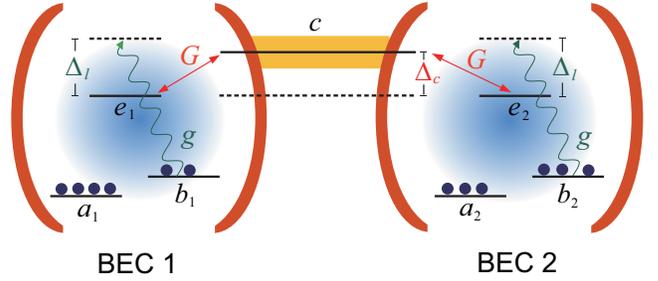}}
   \caption{\label{scheme} (Color online) Scheme for entanglement generation between two spinor Bose-Einstein condensates (BECs). 
    The BECs are placed in an optical cavity, in the strong coupling regime allowing for coherent transfer with coupling $ G $ between a cavity photon and an optical transition between two internal states ($b_{1,2} \leftrightarrow e_{1,2} $) of the atoms.  The BEC exists initially in a spin coherent states between the ground states of the atoms $ a_{1,2}, b_{1,2} $. The two BEC entanglement is initiated with an laser transition $ g $ detuned from the resonance by energy $ \Delta_l $.  The cavities are also detuned from the excitation by an energy $ \Delta_c $. The cavities are connected by an optical fiber forming a common mode $ c $.  }
\end{figure}

The generation of entanglement between BECs is of interest from several perspectives.  Apart from the intrinsic fundamental interest, it allows for performing quantum information processing based on spinor BECs. Recently a novel scheme for performing quantum information processing was introduced based on spin coherent states \cite{byrnes12}.  Existing schemes using spin coherent states are generally based
on the continuous variables approximation, where the total spin is initially polarized in one direction (say $ S^x =N$), then the other total spin variables are used as quasi-position and momentum variables $ x \approx S^y/\sqrt{N}, p \approx S^z/\sqrt{N} $ with $ [x,p]= i $.  In contrast to this approach, in Ref. \cite{byrnes12}, states that are beyond the continuous variable approximation are used to encode the quantum information. The basic idea is to encode a qubit state $\alpha \left | 0 \right \rangle + \beta \left | 1 \right
\rangle$ on the spin coherent state \cite{gross12} 
\begin{align}
   \label{beceq}
   \left |\alpha,\beta \right \rangle \rangle\equiv\frac{1}{\sqrt{N!}}(\alpha
   a^\dagger+\beta
b^\dagger)^{N}|0\rangle
\end{align}
where $a^\dagger$ and $b^\dagger$ are the creation operators of two spin states of the BEC and obeying the commutation relations $\left[a,a^\dagger\right ] =\left [ b,b^\dagger \right ]=1$. The number of atoms in the BEC is $ N $. The properties of entangling two spinor BECs under a $ S^z S^z $ gate have been discussed in detail in Ref. \cite{byrnes13} where it was shown that entanglement shows a ``devil's crevasse'' structure and the states should be robust even in the presence of decoherence for sufficiently short gate times. Using a combination of single spin gates such as that realized by microwave pulses \cite{bohi09,riedel10} and $ S^z S^z $ gates it is possible to perform several types of quantum algorithms \cite{byrnes12,byrnes13}.   

Several schemes have been proposed to generate entanglement between two BECs. The first scheme for generating entanglement
involves using state-dependent forces for two BECs in close proximity \cite{treutlein06}. In this scheme, when state-dependent forces are applied to the BECs, one of the spin components from each BEC overlaps in real space. Due to the $s$-wave scattering between the overlapping atoms, this creates a  $ S^z S^z $ interaction, where $ S^z $ is the total spin in the $z$-component of the BEC. A second
scheme involves generating a geometric phase by applying a common laser mode to two BECs \cite{hussain14}.  A detuned laser induces an ac Stark shift to the BECs, producing entanglement between the mode and the BECs.  By displacing the optical mode in a closed trajectory, the BECs pick up an entangling phase, which can again be written as an $ S^z S^z $ interaction.  A third scheme, which will be studied in detail in this paper, involves coherently exchanging photons between the two BECs by placing them in a cavity \cite{pyrkov13}.  The strong coupling of BECs in optical cavities was demonstrated in Ref. \cite{colombe07}, thus is a potentially experimentally viable method that has the advantage of scalability to a large number BECs. 

While Ref. \cite{pyrkov13} introduced the general theory for the entanglement generation, the effective decoherence rates were estimated based on simulations of subsystems of the scheme, extrapolating this to the whole scheme. In order to gauge the effectiveness of the scheme it is instructive to perform a direct numerical simulation to first confirm that the scheme indeed performs as predicted, and discuss shortcomings under realistic experimental parameters. We find that the scheme works very effectively to create entanglement for the short timescales $ \Omega t \sim 1/N $, corresponding to the continuous variables regime. For intermediate timescales $ \Omega t \sim 1/\sqrt{N} $, a more careful choice of parameters is required in the presence of decoherence effects.  For longer times, $ \Omega t \sim 1 $, the scheme suffers enhanced decoherence effects due to the appearance of Schrodinger cat-like states.  This is in agreement with the prior analysis of \cite{byrnes13} based on a generic dephasing decoherence. 

This paper is organized as follows.  In Sec. \ref{sec:entgen} we review the scheme for entanglement generation using cavity QED, and provide an alternate derivation of the effective $ S^z S^z $ interaction.  In Sec. \ref{sec:num} we discuss the numerical approach that is performed in this study, and the quantities examined. In Sec. \ref{sec:results} the results of our simulations are shown, showing entanglement generation in time, the scaling of entanglement toward large-scale systems, and partial Q-distributions. In Sec. \ref{sec:exp} we estimate relevant experimental parameters and summarize our findings in Sec. \ref{sec:conc}.

\section{Entanglement generation}
\label{sec:entgen}

We now give a brief description of the entangling scheme. Our basic setup consists of two BECs placed in an optical cavity as shown in Figure \ref{scheme} \cite{pellizzari97,pyrkov13}. Here $ a_i,b_i $ denote annihilation operators for bosons in the ground state labeled by the two BECs $ i = 1,2 $.  For $ ^{87} \mbox{Rb} $ the two states
would be typically the hyperfine states $ | F =1, m_F=-1 \rangle $ and $ | F = 2, m_F=1 \rangle $ respectively \cite{treutlein06,bohi09,riedel10}. $ e_i $ is an excited state which optically couples to the $ b_i $ state.  Such cavities coupled to BECs have been realized in atom chip systems where strong coupling has been achieved \cite{colombe07}.  The purpose of the cavity is to coherently convert an excitation $ e_i $ into a cavity photon $ p_i $.  The cavities are coupled to each other via an optic fiber, described by the Hamiltonian
\begin{align}
H_f = & \nu ( p_2^\dagger p + p^\dagger p_2 + p^\dagger p_1 + p_1^\dagger p)   \nonumber \\
& + \hbar \omega ( p_1^\dagger p_1 + p_2^\dagger p_2 + p^\dagger p )
\label{fiberham}
\end{align}
where $ p $ is the mode within the fiber, $ \nu $ is the coupling between the cavity and the fiber, and $ \hbar \omega  $ is the energy of the modes. Diagonalizing (\ref{fiberham}), an eigenstate with energy $ \hbar \omega  $ exists of the form $ c = (p_1 - p_2)/\sqrt{2} $, which we use as the common mode linking the two BECs \cite{pyrkov13}.   Alternatively, the BECs may be placed within the same optical cavity,  such that they access the same physical mode. The basic concept of the scheme  (see Fig. \ref{scheme}) is to create an effective $ S^z S^z $ interaction by taking advantage of the following fourth-order process 1) a photon is absorbed by BEC 1 from the laser; 2) the photon is re-emitted through the cavity mode; 3) the photon is absorbed by BEC 2; 4) stimulated transition of the atom back to the ground state by the laser.   

The Hamiltonian coupling the BECs to the common mode can be described \cite{pyrkov13}
\begin{align}
   \label{heff}
   H_c= \Delta_c c^\dagger c + \sum_{i=1,2} G(e_i^\dagger b_i c + b_i^\dagger e_i c^\dagger)
\end{align}
where $ \Delta_c $ is the detuning between the cavity and the $ b \leftrightarrow e $ transition and  $ G $ is the atom-cavity mode coupling.  In addition, we have a controllable laser field with the Hamiltonian
\begin{align}
H_l= \sum_{i=1,2}g( e_i^\dagger b_i + b_i^\dagger e_i ) + \Delta_l e^\dagger_i e_i 
\label{laserham}
\end{align}
where $ \Delta_l $ is the detuning between the laser transition and the $ b \leftrightarrow e $ transition and $ g $ is the laser coupling.   

The above Hamiltonian can lead to an entangling gate as described in Ref. \cite{pyrkov13}. This can be seen by adiabatically eliminating the photon mode $ c $ by setting $ \frac{dc}{dt}=0 $, then eliminating the excited state $ e_i $ by setting $ \frac{d e_i}{dt}=0 $, giving the effective Hamiltonian 
\begin{align}
H_\text{\tiny ad} & = - \frac{ 2 G^2 g^2}{\Delta_c \Delta_l^2} ( b^\dagger_1 b_1 + b^\dagger_2 b_2)( b^\dagger_1 b_1 + b^\dagger_2 b_2) 
- \frac{g^2}{\Delta_l} ( b^\dagger_1 b_1 + b^\dagger_2 b_2) .
\end{align}
The first term is a fourth order interaction which will give rise to an effective interaction between the BECs.  The second term is the ac Stark shift term due to the laser (\ref{laserham}). This may be rewritten in terms of spin operators $  S^z_i = a^\dagger_i a_i - b^\dagger_i b_i $ assuming that the total number on each BEC is fixed and equal $ N = a^\dagger_i a_i + b^\dagger_i b_i $ as 
\begin{align}
H_\text{\tiny eff} & =\hbar \omega (S^z_1+ S^z_2) -\hbar \Omega \left[S^z_1 S^z_2 +  \frac{(S^z_1)^2}{2} + \frac{(S^z_2)^2}{2}\right] 
\label{effham}
\end{align}
where we have dropped constant terms and defined
\begin{align}
\hbar \omega & = \frac{g^2}{2 \Delta_l} + \frac{G^2 g^2 N }{\Delta_c \Delta_l^2} , \label{singlerot} \\
\hbar \Omega & = \frac{G^2 g^2}{2 \Delta_c \Delta_l^2} .
\label{hbaromega}
\end{align}
We thus see that in addition to a rotational term $ \omega $, there is an entangling interaction $ S^z_1 S^z_2 $ that is produced by the above Hamiltonian. The procedure also produces a self-interaction Hamiltonian $ (S^z_i)^2 $ on each of the BECs, which corresponds to a squeezing term \cite{gross12,riedel10}. The squeezing term is an unwanted by-product of the scheme for our purposes, and is present in all schemes as listed in the introduction \cite{treutlein06,hussain14,pyrkov13}.  Methods to remove the squeezing term have been discussed in 
Ref. \cite{hussain14}, where a two-step process is used to first create the desired $ S^z_1 S^z_2 $ entanglement, then each BEC is squeezed in the reverse direction to cancel off these effects.

\section{Numerical methods}
\label{sec:num}

To analyze the performance of the scheme, we consider the following master equation taking in to account of cavity loss and spontaneous emission
\begin{align}
   \frac{d\rho}{dt} = &  -\frac{i}{\hbar} \left[H_c+ H_l,\rho\right]+
	\frac{\Gamma_s}{2}
   \sum_{i=1,2} (\mathcal{D}[F_{ai}^{-}]\rho + \mathcal{D}[F_{bi}^{-}]\rho ) \nonumber \\
	& +\frac{\Gamma_c}{2}    \mathcal{D}[c]\rho ,	   \label{masteqdec}
\end{align}
where $\mathcal{D}[\hat{o}]\rho\equiv 2 \hat{o} \rho \hat{o}^\dagger -\hat{o}^\dagger \hat{o} \rho - \rho \hat{o} \hat{o}^\dagger$, $F_{ai}^- \equiv a_i^\dagger e_i$, $F_{bi}^- \equiv b_i^\dagger e_i$. The first term describes the coherent dynamics of the system, as discussed in the previous section.  The second term describes spontaneous emission, where we assume that the excited state $ e_i $ decays to both levels $ a_i $ and $ b_i $ at the same rate  $\Gamma_s$ for simplicity.  The third term describes cavity decay due to leakage of the cavity photon through the mirrors at a rate $ \Gamma_c$.  

The density matrix is expanded in the Fock basis of the atoms and the photons
\begin{align}
\rho = \sum_{\substack{k_1n_1k_2n_2l \\ k_1'n_1'k_2'n_2'l'}} \rho_{k_1n_1k_2n_2lk_1'n_1'k_2'n_2'l'} |k_1'n_1'k_2'n_2'l
\rangle  \langle k_1n_1k_2n_2l|
\label{rhoexpansion}
\end{align}
where $ k_i $ is the number of atoms in level $ b_i $, $ n_i $ is the number of atoms in level $ e_i $, and $ l $ is the number of photons in the cavity mode. The number of atoms in level $ a_i $ is $N-k_i-n_i$. To solve (\ref{masteqdec}), we use the backwards Euler method to numerically evolve each matrix element in time \cite{butcher08}. A cutoff is imposed on $ n_i $ and $ l $ which is valid under highly detuned conditions that we are interested in.  For all our calculations $ n_i, l \in [0,1] $ was sufficient to maintain $ \mbox{Tr} \rho = 1 $ to a very good approximation throughout the simulation.  

For a given simulation time, we may estimate the amount of entanglement generated by the scheme by calculating the logarithmic 
negativity \cite{peres96,vidal02}. Starting from the expansion (\ref{rhoexpansion}) we take the partial trace to obtain the ground state density matrix
\begin{align}
\tilde{\rho} = \mbox{Tr}_{n_1 n_2 l} [ \rho]  = \sum_{n_1 n_2 l} \langle n_1 n_2 l| \rho | n_1 n_2 l \rangle . 
\end{align}
The entanglement between the two BECs may then be calculated according to 
\begin{align}
   \label{logneg}
   E=\text{log}_{2} ||\tilde{\rho}^{T_1}||=\text{log}_{2} \sum_i | \lambda_i |
\end{align}
where $\rho^{T_1}$ is the partial transpose with respect to $ k_1 $ and $ k_1' $ and $\lambda_i$ are the eigenvalues of $\tilde{\rho}^{T_1}$. We normalize our plots relative to the maximal entanglement possible between the two spinor BECs, given by the state
\begin{align}
\frac{1}{\sqrt{N+1}} \sum_{k=0}^{N} | k \rangle | k \rangle 
\label{maxentstate}
\end{align}
which takes a value
\begin{align}
E_{\mbox{\tiny max}} = \text{log}_2 (N+1) .
\label{emax}
\end{align}
This has the same value for both the logarithmic negativity and von Neumann entropy, valid for pure states.

\section{Results}
\label{sec:results}

\subsection{Entanglement generation}

\begin{figure}
\scalebox{0.4}{\includegraphics{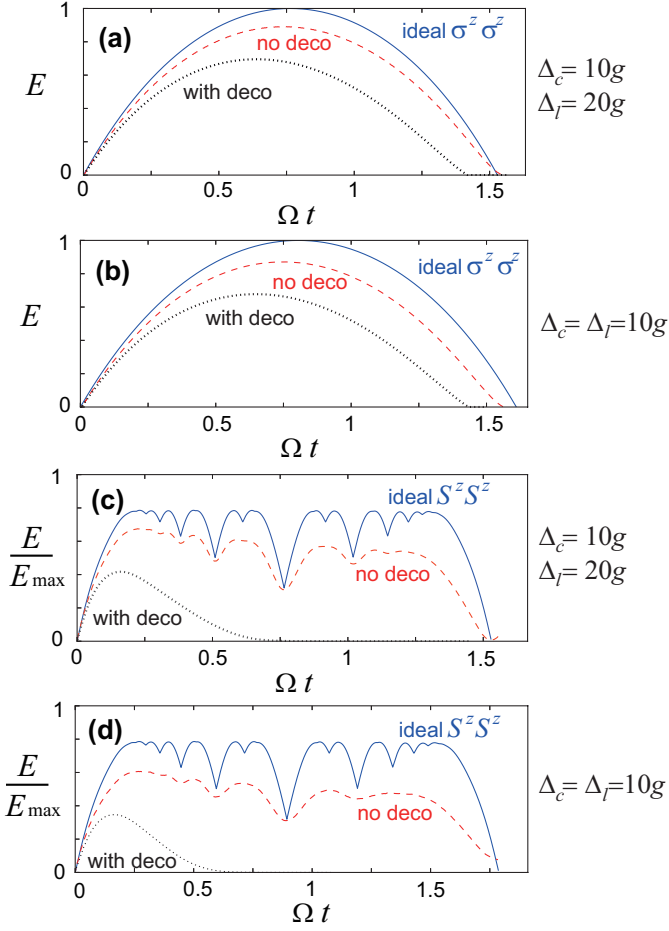}}
\caption{\label{fig2entvt} (Color online)  The entanglement generated by the proposed scheme for (a) (b) $ N = 1 $ and (c) (d) $ N = 8 $. Plots (a) (c) show the off-resonant case with $ \Delta_c/g =  10 $ and $ \Delta_l/g = 20 $ while plots (b) (d) show the resonant case with $ \Delta_c/g =  \Delta_l/g = 10 $.  In each plot the entanglement due to a pure $S^z S^z$ interaction (solid lines), the proposed scheme with $ \Gamma_s = \Gamma_c = 0 $ (dashed lines), and the proposed scheme with cavity decay $\hbar \Gamma_c/g=0.1$ and spontaneous emission $\hbar \Gamma_s /g=0.01$ are shown (dotted lines).  $ G/g = 1 $ is used for all cases and the timescale is measured in units of $ \Omega = \frac{G^2 g^2}{2\hbar \Delta_c \Delta_l^2} $. The timescale of the ideal $S^z S^z$ interaction curves (solid lines) have been adjusted to fit the shape of the simulated curves. }
\end{figure}

Figure \ref{fig2entvt} shows the entanglement as measured by the logarithmic negativity (\ref{logneg}) by numerically evolving (\ref{masteqdec}) from the initial condition $ \rho(t=0) = | \psi(t=0) \rangle \langle \psi(t=0) | $ where 
\begin{align}
| \psi(t=0) \rangle = | \frac{1}{\sqrt{2}},\frac{1}{\sqrt{2}}\rangle \rangle_1  | \frac{1}{\sqrt{2}},\frac{1}{\sqrt{2}}\rangle \rangle_2 
\end{align}
As discussed in Ref. \cite{byrnes13}, for a $ S^z S^z $ interaction the maximal entanglement is generated for states starting in states lying on the equator of the Bloch sphere. From the simulations we can see that our protocol in all cases successfully creates entanglement between the two BECs, even when taking in to account decoherence effects. A non-zero value of the logarithmic negativity is sufficient to show that entanglement is present.  A zero value of logarithmic negativity does not however necessarily mean that entanglement is not present \cite{vidal02}.

To confirm that our simulation correctly reproduces known results, we first consider the $ N = 1 $ case corresponding to a standard qubit.  Setting $ N = 1 $ reduces all spin operators to standard Pauli operators $S^{x,y,z} \rightarrow \sigma^{x,y,z} $, thus our scheme should produce a $ \sigma^z \sigma^z $ two qubit interaction.  To illustrate the performance of the protocol under a variety of conditions, we consider two primary cases of where the intermediate states are  off-resonant ($ \Delta_c \ne \Delta_l $) and resonant ($ \Delta_c = \Delta_l $).  Figure \ref{fig2entvt}(a)(b) shows our results these cases respectively.  We see that without spontaneous emission and cavity decay the entanglement agrees well with the ideal $ \sigma^z \sigma^z $ curve, reaching the maximally entangled state halfway through the entangling operation.  Both the off-resonant and resonant cases produces nearly identical results in terms of the entanglement generated.  We note that the entanglement curves can be made to approach to ideal cases simply by increasing the detuning $ \Delta_{c,l}/g $ to larger values. The discrepancy in the zero decoherence cases to the ideal curves may be attributed to an effective decoherence from tracing out the excited and photonic degrees of freedom, which become populated as a result of the protocol.  For cases including spontaneous decay and cavity photon loss, the entanglement generated is reduced as expected. Our chosen parameters for the spontaneous emission follow the typical experimental parameters (see Sec. \ref{sec:exp}) as set by the cavity as realized in Ref. \cite{colombe07}.  These effects may also be reduced by increasing the detuning, at the expense of producing a slower entangling times according to (\ref{hbaromega}).  

We now turn to the $ N> 1 $ case, as shown in Fig. \ref{fig2entvt}(c)(d).  Due to the exponential growth of the Hilbert space with $ N $, we are restricted in our numerical simulations to relatively small boson numbers $ N\le 8 $.  Nevertheless, our numerical results confirm the expected behavior of the protocol.  As discussed in detail in Refs. \cite{byrnes13,sinatrapaper}, an $ S^z S^z $ gate creates a characteristic ``devil's crevasse'' structure in the entanglement as shown by the solid curves in Figs. \ref{fig2entvt}(c)(d).  Dips in the entanglement originate from resonances in the arrangement of spin coherent states, occurring at times that are a rational multiple of the characteristic timescale. The overall timescale in all cases match to reasonable accuracy returning to an unentangled state at $ \Omega t \approx \pi/2 $.   The number of these dips increase with $ N $, although many of these are seen already in the $ N = 8 $ simulations shown. The amount of entanglement created by the $ S^z S^z $ interaction does not reach the maximal entanglement $ E = E_{\mbox{\tiny max}} $, as a $ S^z S^z $ interaction never creates the state (\ref{maxentstate}) but asymptotically approaches  $ E = E_{\mbox{\tiny max}}/2 $ for large $ N$ \cite{byrnes13}.  For our chosen parameters, the protocol generally gives a similar form to the ideal $ S^z S^z $ curve, but with a reduced amount of entanglement.  Most of the discrepancy occurs away from the dips in the entanglement, with some of the values in the entanglement agreeing very well in the dipped points.  We speculate that this may occur due to the effective decoherence being of a $ S^z $-dephasing form, which produces a similar type reduction in entanglement as seen in Ref. \cite{byrnes13}.

When spontaneous decay and cavity loss are included the entanglement drops significantly, more than that seen in the $ N = 1 $ case.  We attribute this to the sensitiveness of the created states to the effective decoherence created by spontaneous decay and cavity loss.  In Ref. \cite{byrnes13}, it was shown that a $ S^x $-dephasing strongly suppresses the formation of states beyond times $ \Omega t \gtrsim 1/\sqrt{N} $ as Schrodinger cat-like states are created for these times.  However, for relatively short timescales such as $ \Omega t \le 1/\sqrt{N} $ it is possible for stable entanglement to be generated.  We thus expect that in a realistic situation only these short timescale gates are producible reliably. In terms of the production of macroscopic entanglement, at times $ \Omega t = 1/\sqrt{2N} $ it was shown in Ref. \cite{byrnes13} that $ E \sim E_{\mbox{\tiny max}}/2 $ can be created formed.  We analyze the stability of these states in the next subsection.

\subsection{Scaling of entanglement towards large boson number $ N $}

For macroscopic objects it is generally expected that quantum effects, including entanglement, are difficult to observe due to the fast decoherence which sets in for large particle number $ N $. However, it is known that under certain circumstances quantum effects may survive in macroscopic objects. For example, experiments such as that performed in Refs. \cite{julsgaard01,krauter13,bao12}  entanglement between macroscopic ensembles of atoms are created and used to realize teleportation. This apparent disparity is resolved by noting that decoherence inherently depends on the type of state that one starts with.  For example, a Schrodinger cat state, which in our system would take the form
\begin{align}
\frac{1}{\sqrt{2}} \left( | \frac{1}{\sqrt{2}},\frac{1}{\sqrt{2}}\rangle \rangle +  | \frac{1}{\sqrt{2}},\frac{1}{\sqrt{2}}\rangle \rangle \right)
\end{align}
decoheres at a timescale proportional to $ \sim \frac{1}{\Gamma N^2} $, while a spin coherent state $ | \alpha, \beta  \rangle \rangle  $ decoheres at a timescale $ \sim \frac{1}{\Gamma} $ \cite{byrnes12}. This state-dependency of the decoherence means that for various types of entangled state, it is important to compare states of the same type, in order to understand whether a given state will survive in the macroscopic limit.  

For the $ S^z S^z $ interaction that our protocol aims to produce, there are several characteristic types of state that are of particular interest.  The first type of state corresponds to evolution times $ \Omega t = \frac{\pi}{4N} $. As discussed in Refs. \cite{byrnes12,byrnes13}, these entangling times are required for the spin coherent state analogue of the CNOT gate, and the amount of entanglement generated is comparable to a single qubit with $ E \sim O(1) $.  In terms of the ``devil's crevasse'' entanglement curve in Fig. \ref{fig2entvt}(c)(d), this corresponds to very early times in the evolution, at the base of the initial turn-on of the entanglement. 
Gate times of the order $ \Omega t \sim 1/N $ only cause a small redistribution of the spin coherent states around the Bloch sphere, thus are still within the continuous variable approximation.  In the presence of dephasing, such states are rather robust and are expected to survive in the macroscopic limit $ N \rightarrow \infty $ \cite{byrnes13}.  The second characteristic state occurs for evolution times $ \Omega t \sim \frac{1}{\sqrt{N}} $.  At these timescales, the spin coherent states are distributed evenly around the equator of the Bloch sphere, each entangled with a $ S^z $ eigenstate on the other BEC.  The amount of entanglement generated at these times is of the order of the maximal entanglement between the BECs, in the sense that $ E \sim O(E_{\mbox{\tiny max}}) $.  In the presence of decoherence, these states decay with a power law with $ N $, thus under suitable conditions can be observable for large scale systems \cite{byrnes13}.  For timescales $ \Omega t > \frac{1}{\sqrt{N}} $, very fragile Schrodinger cat-like states are generated, which exponentially decay in the presence of 
decoherence \cite{byrnes13}.   For a more detailed description of the types of states that are created, we refer the reader to Ref. \cite{byrnes13}, in which we investigate the effects of dephasing on the entanglement generated by a pure $ S^z S^z $ interaction. We summarize the general behavior according to
\begin{align}
   \frac{E(\Gamma)}{E(\Gamma=0)} = \left\{
     \begin{array}{lr}
        \text{const.} & \Omega t \sim 1/N\\
        N^{-\gamma}& \Omega  t \sim 1/\sqrt{N}\\
        e^{-N^2} & \Omega  t \sim O(1)
     \end{array}
   \right. 
\end{align}
where $\gamma$ is a parameter such that $0<\gamma<1$. Therefore under realistic conditions we expect that only states $ \Omega t \le \frac{1}{\sqrt{N}} $ are observable for large systems. Even for the relatively small systems simulated in this paper, the results of Fig. \ref{fig2entvt}(c)(d) confirm this expectation, where the entanglement quickly decays to zero for these timescales. For these reasons henceforth we examine the two timescales $ \Omega t = \frac{\pi}{4N},\frac{1}{\sqrt{N}} $.

\begin{figure}
\scalebox{0.45}{\includegraphics{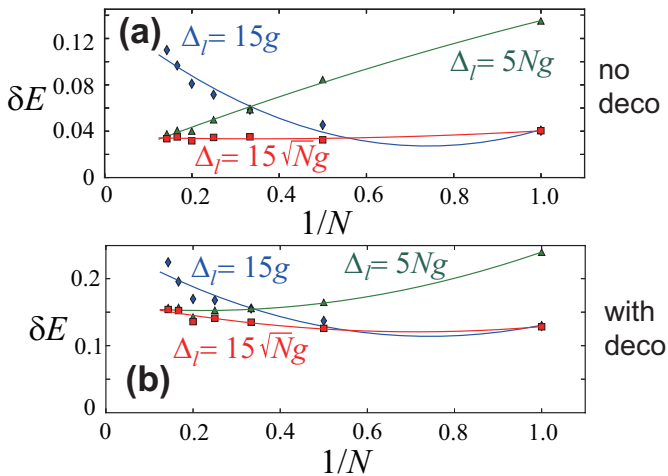}}
   \caption{\label{fig4scalingsqrt2n} (Color online)  Entanglement at the time $ \Omega t = \frac{1}{\sqrt{2N}} $ for various boson numbers $ N $.  $ \delta E $ is the difference between the ideal entanglement generated by a pure $ S^z S^z $ interaction and the proposed scheme. Once again, each plot shows the scaling of three different possible detunings: $\Delta_l/g=15$ (diamonds), $\Delta_l/g=15\sqrt{N}$ (squares), and $\Delta_l/g=5N$ (triangles). The proposed scheme with $ \Gamma_s = \Gamma_c = 0 $ (a) and the proposed scheme with cavity decay $\hbar \Gamma_c/g=0.1$ and spontaneous emission $\hbar \Gamma_s /g=0.01$ are shown (b).  $ G/g = 1 $, $ \Delta_c/g=2 $  is used for all cases.} 
\end{figure}

Figure \ref{fig4scalingsqrt2n} shows the entanglement scaling for the difference $ \delta E $ for the time $ \Omega t = \frac{1}{\sqrt{N}} $.  We plot the difference $ \delta E $ between the normalized entanglement $ E/E_{\mbox{\tiny max}} $ generated by an ideal $ S^z S^z $ interaction and the proposed scheme as a measure of how well the entanglement survives in the macroscopic limit $ 1/N \rightarrow 0 $. 
Looking at the case of constant detuning $ \Delta_l/g = 15 $ (Fig. \ref{fig4scalingsqrt2n}(a)), we see poor scaling behavior in both the decoherence free case and including spontaneous emission and cavity decay, which may suggest that the protocol fails for large systems. The reason for this failure can be attributed to the effective decoherence due to spontaneous decay being enhanced due to the atom number $ N $. As discussed in Ref. \cite{pyrkov13}, the spontaneous decay creates a decoherence at the effective rate
\begin{align}
\Gamma_s^{\mbox{\tiny eff}} = \frac{\Gamma_s g^2 N}{\Delta_l^2} .
\label{effdec}
\end{align}
We may thus compensate for the enhanced spontaneous emission by choosing increasingly large detunings as $ N $ is increased. This suggests that a suitable strategy may be to choose the detuning  $ \Delta_l \propto  \sqrt{N} $ such as to cancel the $ N $ dependence in (\ref{effdec}).  For comparison we also try the more aggressive detuning strategy $ \Delta_l \propto N $.  

Using the scaling detunings, we now see that $ \delta E $ behaves more favorably in both the cases with and without decoherence as $ N $ is increased (Fig. \ref{fig4scalingsqrt2n}).  These results would suggest that in order to generate states of the type $ \Omega t = \frac{1}{\sqrt{N}} $ it is important to choose a detuning that is sufficiently large to overcome the spontaneous decay effects, which have a stronger effect to destroy this class of state.  We note that for the dephasing case previously studied in Ref. \cite{byrnes13}, it was found that the entanglement should scale as a power law for $ \Omega t = \frac{1}{\sqrt{N}} $, thus although no entanglement survives in the limit of $ N \rightarrow \infty $, for large but finite systems some entanglement can survive.

\begin{figure}
\scalebox{0.45}{\includegraphics{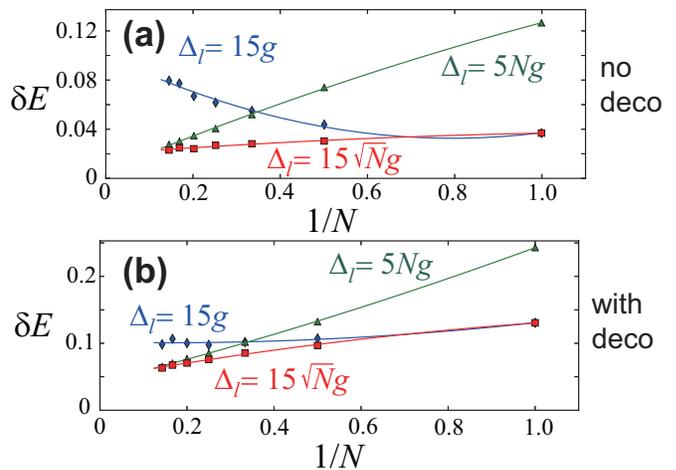}}
   \caption{\label{fig3scalingpi4N} (Color online) Entanglement at the time $ \Omega t = \frac{\pi}{4N} $ for various boson numbers $ N $.  $ \delta E $ is the difference between the ideal entanglement generated by a pure $ S^z S^z $ interaction and the proposed scheme. Each plot shows the scaling of three different possible detunings: $\Delta_l=15$ (diamonds), $\Delta_l=15\sqrt{N}$ (squares), and $\Delta_l=5N$ (triangles). The proposed scheme with $ \Gamma_s = \Gamma_c = 0 $ (a) and the proposed scheme with cavity decay $\hbar \Gamma_c/g=0.1$ and spontaneous emission $\hbar \Gamma_s /g=0.01$ are shown (b).  $ G/g = 1 $, $ \Delta_c/g=2 $ is used for all cases. 
}
\end{figure} 

Figure \ref{fig3scalingpi4N} shows the scaling of the entanglement with boson number $ N $ at the time $\Omega t = \frac{\pi}{4N} $. 
We see that states created around this characteristic time show good scaling with increasing $N$, becoming increasingly closer to the pure interaction. The cases including spontaneous decay and cavity photon loss show a strong improvement as the system size is increased. This can be attributed to the shorter physical gate times $\Omega t = \frac{\pi}{4N} $ as the $ N $ is increased.  The shorter times that the system are evolved mean that there is less time for spontaneous decay and cavity photon loss to occur, which improve the performance of the scheme. The same general behavior was seen in the scaling in Ref. \cite{byrnes13} for an $ S^z S^z $ interaction in the presence of dephasing.  These calculations suggest that states with $\Omega t = \frac{\pi}{4N} $ should be able to be robustly created for macroscopic systems.

\subsection{Partial Q-distributions}

To visualize the states that are generated by the current protocol, it is instructive to plot the Q-distribution to show the distribution of the states on the Bloch sphere \cite{gross12}.  For an ideal $ S^z S^z $ interaction, the state that is generated is written \cite{byrnes13}
\begin{align}
& e^{-i \Omega S^z_1 S^z_2 t} | \frac{1}{\sqrt{2}}, \frac{1}{\sqrt{2}} \rangle \rangle_1 
| \frac{1}{\sqrt{2}}, \frac{1}{\sqrt{2}} \rangle \rangle_2 \nonumber \\
& = \frac{1}{\sqrt{2^N}} \sum_k  \sqrt{N \choose k} | \frac{e^{i(N-2k) \Omega t}}{\sqrt{2}} , \frac{e^{-i(N-2k) \Omega t}}{\sqrt{2}} \rangle \rangle_1 | k \rangle_2 .
\label{entangledstate}
\end{align}
As discussed in detail in Ref. \cite{byrnes13}, we may understand this state as an entangled state between number states on BEC 2 and spin coherent states distributed at various locations on the equator of the Bloch sphere on BEC 1. We note that there is no asymmetry between BECs 1 and 2 as the above expression may be rewritten by expanding BEC 1 in the number basis and leaving BEC 2 in the spin coherent state basis.  We may create visualizations of the state by projecting the entangled state on number states of BEC 2, then calculating the partial Q-distributions for BEC 1 defined as 
\begin{align}
Q_{k_2} (\alpha,\beta) = \frac{N+1}{4 \pi} \langle \langle \alpha, \beta | \langle k_2 |\tilde{\rho} | k_2 \rangle | \alpha, \beta \rangle  \rangle  .
\label{qfuncdef}
\end{align}
Due to the projection on the  $ | k_2 \rangle  $ basis, the integral of the distributions (\ref{qfuncdef}) are no longer normalized to unity.  The integrated distribution gives the probability of the particular $ | k_2 \rangle $-coherent state pairing, which depends on the amplitude of the term in (\ref{entangledstate}). 

\begin{figure}
\scalebox{0.42}{\includegraphics{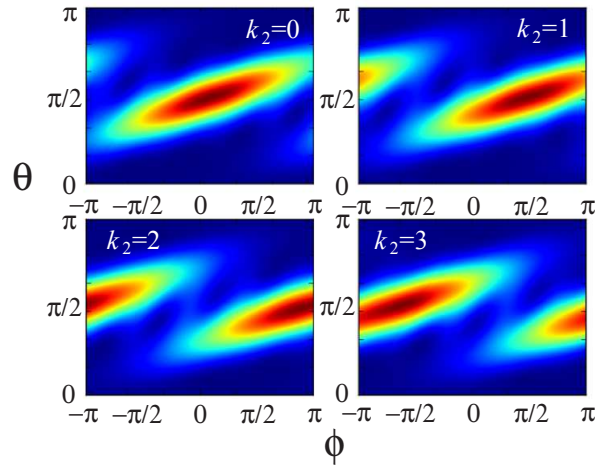}}
   \caption{\label{fig5qfuncs} (Color online)  A plot of the partial Q-distribution of BEC 1 at time
      $ \Omega t=1/\sqrt{2N}$.  Figures show (\ref{qfuncdef}) with projections on BEC 2 for the $ k_2 $ values shown.  Parameters used are $ \Delta_c/g = 2$, $\Delta_l/g = 20 $, $ G/g = 1 $, $ N = 8 $.}
\end{figure}

Figure \ref{fig5qfuncs} shows the partial Q-distributions for a state generated using the off-resonant scheme for the state $ \Omega t = 1/\sqrt{2N} $.  As expected, we see that for various $ k_2 $ projections, the states on BEC 1 are distributed at various positions on the equator of the Bloch sphere, in accordance with (\ref{entangledstate}).  One difference to (\ref{entangledstate}) is that instead of spin coherent states on BEC 1, which would appear as symmetric Gaussian distributions for the Q-distribution, we have diagonally squeezed states.  This may be attributed to the self-interaction term $ (S^z_1)^2 $ that are present in (\ref{effham}) which produce a squeezing of the distribution \cite{gross12,riedel10}.  If one wishes to generate a pure $ S^z S^z $ interaction, it is necessary to cancel the squeezing terms in the Hamiltonian, which may be achieved by methods such as that discussed in Refs. \cite{pyrkov13}, where the detuning is reversed to produced a flip in the sign of the effective interaction (\ref{hbaromega}).  

\begin{figure}
\scalebox{0.42}{\includegraphics{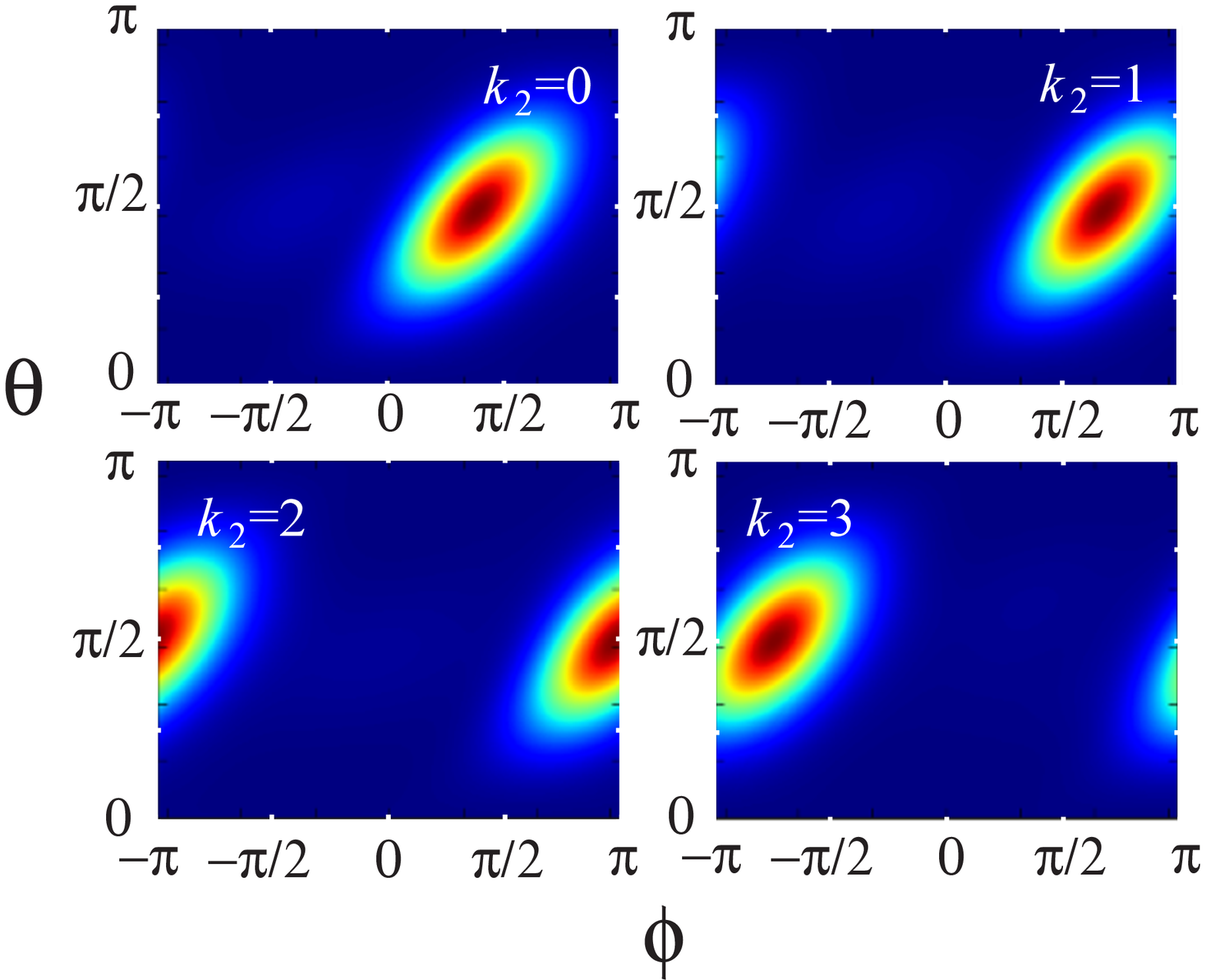}}
   \caption{\label{fig6qfuncs} (Color online)  A plot of the partial Q-distribution of BEC 1 at time $\Omega t=1/\sqrt{2N}$.  Figures show (\ref{qfuncdef}) with projections on BEC 2 for the $ k_2 $ values shown.  Parameters used are $ \Delta_c/g = \Delta_l/g = 10 $, $ G/g = 1 $, $ N = 8 $.  }
\end{figure}

Figure \ref{fig6qfuncs} shows the partial Q-distributions using the resonant scheme for the same time.  The plots show the  Gaussian distributions distributed at various equatorial positions displaying the expected $ S^z S^z $ correlations.  Each of the distributions show unexpectedly far less squeezing in comparison to Fig. \ref{fig5qfuncs}.  This may be explained by the presence of spurious anti-squeezing terms introduced by the $ n_i \in [0,1] $ cutoff imposed in the simulations.  For a state with $ k_i $ atoms in level $ b_i $, the The Hamiltonian (\ref{laserham}) 
produces an ac Stark shift that can be written exactly
\begin{align}
\delta \epsilon = \frac{k_i}{2} (\Delta_l - \sqrt{\Delta_l^2 + 4g^2}).  
\end{align}
This may be written in terms of spin operators as 
\begin{align}
H_{\mbox{\tiny ac}} = \frac{1}{4} (\sqrt{\Delta_l^2 + 4g^2}-\Delta_l) (S^z_1 + S^z_2)
\label{fullacstark}
\end{align}
where we have dropped constant terms.  Assuming $ \Delta_l \gg g $ we obtain the first term in (\ref{singlerot}).  While it is clear from (\ref{fullacstark}) that no squeezing should result from  (\ref{laserham}), the truncation introduced in the simulation artificially introduces this.  Assuming a maximum of one excited state in $ e_i $, the Hamiltonian (\ref{laserham}) becomes
\begin{align}
H_l \approx \left(
\begin{array}{cc}
0 & g\sqrt{k_i} \\
g\sqrt{k_i} & \Delta_l 
\end{array}
\right) .
\end{align}
The ground state correction is then
\begin{align}
\delta \epsilon_{\mbox{\tiny spurious}} = \frac{1}{2} (\Delta_l - \sqrt{\Delta_l^2 + 4 g^2 k_i}).  
\end{align}
which on expanding the square root apparently produces an effective Hamiltonian
\begin{align}
H_{\mbox{\tiny spurious}} = \frac{g^2}{2 \Delta_l}  (S^z_1 + S^z_2) +  \frac{g^4}{4 \Delta_l^3} ((S^z_1)^2 + (S^z_2)^2) + \dots .
\label{spuriousham}
\end{align}
While the ac Stark shift is correctly predicted in the first term of (\ref{spuriousham}), a spurious spin squeezing term is created due to the truncation.  For the parameters chosen in Fig.  \ref{fig6qfuncs} this numerically cancels the squeezing terms in (\ref{effham}). For the parameters chosen in Fig. \ref{fig5qfuncs}, the squeezing terms in (\ref{effham}) are much larger than the spurious anti-squeezing terms, hence we expect that the realistic distributions will more resemble those shown in Fig. \ref{fig5qfuncs}.

\section{Estimated experimental parameters}
\label{sec:exp}

We now estimate the physical gate times that are expected for the proposed scheme.  Using numbers given in Ref. \cite{colombe07}, we have $G/\hbar = 2\pi \times 215 \mbox{MHz}, \Gamma_s = 2\pi \times 3 \mbox{MHz}, \Gamma_c = 2\pi \times 53 \mbox{MHz} $. We assume that the BEC particle number is $ N = 10^3 $ and $ g = G $. As discussed above, it is important to scale the detunings with the particle number in order to suppress decoherence induced by the spontaneous emission and the cavity decay. Choosing $ \Delta_c = 2g $ and $ \Delta_l = 15\sqrt{N} g $, 
the physical time required to generate a $ \Omega t = \frac{\pi}{4N} $ state is
\begin{align}
t = \frac{\pi}{4 N \Omega } = 520 \mbox{ns}
\end{align}
For the $ \Omega t = \frac{1}{\sqrt{2N}} $ state, the physical gate time using the same parameters is
\begin{align}
t = \frac{1}{\sqrt{2N} \Omega } = 15 \mu \mbox{s} .
\end{align}
These should be compared to the effective decoherence time due to spontaneous emission (\ref{effdec}) giving $ 1/ \Gamma_s^{\mbox{\tiny eff}} = 12 \mu \mbox{s} $. The effective decoherence due to cavity loss is \cite{pyrkov13}
\begin{align}
\Gamma_c^{\mbox{\tiny eff}} = \frac{\Gamma_c g^2}{\Delta_l^2}
\end{align}
which gives a value $ 1/\Gamma_c^{\mbox{\tiny eff}} = 680 \mu \mbox{s} $.  The above parameters show that generating a $ \Omega t = \frac{\pi}{4N} $ state should be well within the capabilities of current technology. The $ \Omega t = \frac{1}{\sqrt{2N}} $ state is more challenging but still is within the scope of realizability.

\section{Summary and conclusions}
\label{sec:conc}

We have studied a protocol to produce $ S^z S^z $ interactions between two spinor BECs based on the coherent exchange of cavity photons between two BECs.  Our numerical simulations show that entanglement between the BECs can be created under realistic conditions including spontaneous emission and cavity decay.  Various types of states classified according to the dimensionless entangling time $ \Omega t $ have different levels of stability with respect to spontaneous emission and cavity decay.  States with $ \Omega t = \frac{\pi}{4N} $ are rather robust and have a favorable scaling for large scale systems, with moderate detunings. States with $ \Omega t = \frac{1}{\sqrt{2N}} $ require larger detunings to counteract the effects of spontaneous decay in particular, and require detunings that scale with the boson number.  For states with $ \Omega t \gtrsim \frac{1}{\sqrt{N}} $, the $ S^z S^z $ interaction produces Schrodinger cat-like states hence are rather fragile in the presence of decoherence.  We thus expect that our scheme will work for the production of states with entangling times $ \Omega t \lesssim \frac{1}{\sqrt{2N}} $, but will be difficult for states beyond these times.  Fortunately, for quantum information applications such as that discussed in Refs. \cite{byrnes12,pyrkov13b} only short timescale gates with $ \Omega t \le \frac{1}{\sqrt{2N}} $ are necessary, hence our scheme should be suitable for such applications.  

Analysis of the generated entangled states by the partial Q-distribution show the expected behavior, with an additional spin squeezing term due to $ (S^z)^2 $ self-interactions.  One of the advantages of the current approach is in its scalability, where potentially many BECs can be entangled with each other.  The use of photons allow for long-distance entanglement to be produced, which can be scaled up to a quantum network \cite{pyrkov13}.  Meanwhile the use of BECs allow for rather stable quantum memory elements, owing to their long coherence times.

\section*{ACKNOWLEDGMENTS}

This work is supported by the Transdisciplinary Research Integration Center, the Okawa foundation, the Inamori foundation, NTT, and JSPS KAKENHI Grant Number 26790061.

%

\end{document}